\documentclass[aps,preprint,showpacs,superscriptaddress]{revtex4}
\usepackage{amssymb}
\usepackage{graphicx}
\newcommand\ket[1]{\ensuremath{|#1\rangle}}

\begin{document}
\title{ Secure quantum key distribution in an easy way}
\author{Jia-Zhong Hu}
\affiliation{Department of Physics, Tsinghua University, Beijing
100084, China}
\author{Xiang-Bin Wang}
\email{xbwang@mail.tsinghua.edu.cn}
\affiliation{Department of
Physics, Tsinghua University, Beijing 100084, China}

\begin{abstract}
Secure quantum key distribution can be achieved with an imperfect
single-photon source through implementing the decoy-state method.
However, security of all those theoretical results of decoy-state
method based on the original framework raised by Hwang needs
monitoring the source state very carefully, because the elementary
proposition that the counting rates of the same state from different
sources are equal does not hold in general when the source is
unstable. Source intensity monitoring greatly decreases the system
efficiency. Here without using Hwang's proposition for stable
source, we present a sufficient condition for a secure decoy-state
protocol without monitoring the source intensity.  The passive
2-intensity protocol proposed by Adachi, Yamamoto, Koashi, and
Imoto(AYKI) (Phys. Rev. Lett. 99, 180503 (2007) ) satisfies the
condition. Therefore, the protocol can always work securely without
monitoring the source state or switching the source intensity. We
also show that our result can greatly improve the key rate of the
3-intensity protocol with a fluctuating coherent-state source.
\end{abstract}


\maketitle


\section{ Introduction}  Imperfect single-photon source is used in
most of the existing set-ups of quantum key distribution
(QKD)\cite{BB84,GRTZ02,ILM}. To make it secure even under
photon-number-splitting attack \cite{PNS,PNS1}, one can use the
decoy-state method \cite{rep, H03, wang05, LMC05,HQph,AYKI,haya}
based on the general theory of QKD with an imperfect source which
was first built by Inamori, L\"utkenhaus and Mayers\cite{ILM} and
then father studied with some improved results\cite{gllp}. There are
also some other methods \cite{scran,kko} which can make secure QKD
with an imperfect source.

Though many experiments on the decoy-state QKD have been done
\cite{peng},  the original decoy-state theory\cite{H03,wang05,LMC05}
assumes perfect control of the source states in the photon number
space. This is an impossible task for any real set-up in practic. A
new non-trivial problem arises in practice is how to carry out the
decoy-state method securely and efficiently with an unstable source.

The most important proposition of the decoy-state method given by
Hwang\cite{H03} is
\begin{equation}
s_k =s_k'\label{elem}
\end{equation}
where $s_k$ and $s_k'$ are the yield (counting rate) of those
$k-$photon pulses from the decoy source and that from the signal
source. However, as pointed out in Ref.\cite{rep,wangyang,yi} with
concrete examples that this elementary proposition does not always
hold for a fluctuating source. (Note that the fluctuation of each
individual pulses are in general not independent.)

To solve this problem, a number of theoretical works have been
done\cite{wang07,wangapl,wangyang,yi,njp,yilo}. In particular,
without using Eq.(\ref{elem}), a general formula was given in
Ref.\cite{wangyang} for the 3-intensity decoy-state method with
whatever type of error pattern in the source, if we can monitor the
range of a few parameters of each individual pulse state out of
Alice's lab. Although source monitoring can be done with the
existing technology by detecting the intensity of a strong father
pulse in the beginning\cite{guo1,guo2}, it will raise the cost of
the real QKD system and decrease the outcome efficiency because the
repetition rate of strong pulse detection is low. Therefore, it is a
meaningful job to find a class of decoy-state protocols which can
work securely without monitoring any source fluctuation .

Here, without using Eq.(\ref{elem}), we shall give the condition
that a decoy state protocol can work securely with a fluctuating
source. We then show that the 2-intensity passive protocol proposed
by  Adachi, Yamamoto, Koashi, and Imoto (AYKI)\cite{AYKI} meets the
condition of our theorem.

In what follows, we shall first derive an improved formula for a
2-intensity decoy-state protocol with source fluctuation. Based on
this, we propose a theorem on the condition for a protocol to work
securely without source monitoring. Directly applying the theorem we
find AYKI protocol meets the condition and the main formula in AYKI
protocol is independent of whatever source intensity fluctuation. We
then extend our formula to the 3-intensity protocol and show its
high efficiency for a coherent-state based protocol with numerical
simulation.

\section{Our theorem  AYKI protocol}
 In a 2-intensity protocol, Alice has 2 (virtual) sources,
the decoy source $Y$ which prepares decoy pulses, and the signal
source Y' which  prepares signal pulses.  At any time $i$, each
source prepares a pulse and Alice randomly choose one of them
sending out to Bob. Suppose at any time $i$ Alice has a probability
$p_i,\;p_i'$ to sends out the pulse from source  $Y$ and $Y'$,
respectively ($p_i+p_i'=1$). Note that $p_i,\;p_i'$ are not
necessarily constant values here.  Denote $\rho_i,\; \rho_i'$ as
state of a decoy pulse and that of a signal pulse produced at the
$i$th time:
\begin{equation}\label{rhoi}\rho_i=\sum_{k=0}^J
a_{ki}|k\rangle\langle k|\; ;\;\rho_i'=\sum_{k=0}^J
a_{ki}'|k\rangle\langle k|.\end{equation}  Here $J$ can be either
finite or infinite.

Definitely, it makes no difference if all  pulses sent out to Bob
are actually produced by only one physical source. But we assume
there are two sources for clarity in presentation.

In the whole protocol, Alice sends Bob $M$ pulses, one by one. In
response to Alice, Bob observes his detector for $M$ times. As Bob's
{\em i}th observed result, Bob's detector can either click or not
click. If the detector clicks in Bob's {\em i}th observation, then
we say that ``the {\em i}th pulse from Alice has caused a count". We
disregard how the {\em i}th pulse may change after it is sent out.
When we say that Alice's {\em i}th pulse has caused a count we
actually mean that pulse is accompanied by a click at Bob's side at
Bob's {\em i}th observation.

Given the source state in Eqs.(\ref{rhoi}), any $i$th pulse sent out
by Alice must be in a specific photon-number state. To anyone
outside Alice's lab, it looks as if that Alice only sends a photon
number state at each time $i$: sometimes it's vacuum, sometimes it's
a single-photon pulse, sometimes it is a $2-$photon pulses, and so
on. We shall make use of this fact that any individual pulse is in
one Fock state. On the other hand, pulses sent out to Bob can also
be classified by which source it comes from, i.e., a decoy pulse if
it is from the decoy source or a signal pulse if it is from the
signal source.

Given Eqs.(\ref{rhoi}), we have the following  formulas for
$\mathcal P_{di|k}$ and $\mathcal P_{si|k}$ , the probabilities that
the {\em i}th pulse comes from the decoy source or signal source; if
the {\em i}th pulse contains {\em k} photons:
\begin{equation}\label{popdk}
 \mathcal P_{di|k}=p_ia_{ki}d_{ki},\;
 \mathcal P_{si|k}=p'_ia_{ki}'d_{ki}
 \end{equation}
 where
 \begin{equation}\label{dki}
d_{ki}=\frac{1}{p_ia_{ki}+p'_ia_{ki}'},\;{\rm for}\; k\ge0.
\end{equation}
Based on these, we can build up equations which lead to lower bound
of the number of single-photon counts.
\subsection{Definitions} We postulate some definitions
first: Set $\Omega=\{i=1,2,\cdots M\}$, it contains all $i$.  Set
$C$ contains any pulse that has caused a count; set $c_k$ contains
any $k-$photon pulse that has caused a count. Mathematically, the
sufficient and necessary condition for  $i\in C$ is that the $i$th
pulse has caused a count. The sufficient and necessary condition for
$i\in c_k$ is that the {\em i}th pulse contains $k$ photons and it
has caused a count. For instance, if the photon number states of the
first 10 pulses from Alice are
$|0\rangle,\;|0\rangle,\;|1\rangle,\;|2\rangle,\;|0\rangle,\;
|1\rangle,\;|3\rangle,\;|2\rangle,\;|1\rangle,\;|0\rangle,\;$  and
the pulses of $i=2,\; 3,\; 5,\; 6,\; 9,\; 10$ each has caused a
count at Bob's side, then we have
\begin{equation}
C=\{i|i=2,\;3,\;5,\;6,\;9,\;10,\cdots\};\;c_0=\{i|i=2,5,10,\cdots\};
\; c_1=\{i|i=3,6,9,\cdots\}.
\end{equation}
 The number of counts caused by $k-$photon pulse
$n_k$ is just the number of elements in set $c_k$. We shall use
notation $n_{kd},n_{ks}$ for the the number of counts caused by a
$k-$photon decoy pulse and a $k-$photon signal pulse, respectively.
These numbers cannot be directly observed in the experiment.
Obviously, $n_{kd}+n_{ks}=n_k$.
Suppose in the whole protocol, there are $N_d$ counts caused by
decoy pulses and $N_s$ counts caused by signal pulses. Note that
$N_d$ and $N_s$ can be directly observed in the protocol therefore
we regard them as {\em known} values.  Our goal is to find a formula
for $n_{1s}$, i.e., to formulate $n_{1s}$ by $n_{0s},\; n_{0d}$ and
the known values $N_d, \;N_s$ and the bound values of the parameters
in the decoy state and signal state of Eq.(\ref{rhoi}). (In a
2-intensity protocol, values of $n_{0d}$ or $n_{0s}$ is unknown, but
one can still calculate the final key rate with worst-case
estimation\cite{LMC05,AYKI}.)
\subsection{Derivation of Main formulas}
With definitions postulated earlier, we have
\begin{equation}\label{popd}
N_d=\sum_{k=0}^J n_{kd},\; N_s=\sum_{k=0}^J n_{ks}
\end{equation}
 Asymptotically,
\begin{equation}\label{ndsk}
n_{kd} =\sum_{i\in c_k}\mathcal P_{di|k};\; n_{ks} =\sum_{i\in
c_k}\mathcal P_{si|k}
\end{equation}
Consequently,
\begin{equation}\label{popde}
N_d= n_{0d}+ \sum_{i\in c_1}p_ia_{1i}d_{1i}+\sum_{k=2}^J\sum_{i\in
c_k}p_i a_{ki}d_{ki},\end{equation}
\begin{equation}\label{popse}
N_s= n_{0s}+\sum_{i\in c_1}p'_ia_{1i}'d_{1i}+\sum_{k=2}^J\sum_{i\in
c_k}p'_ia_{ki}'d_{1i}.\end{equation}

We want to find the lower bound of value $n_{1s}$.
 Recall the
definition of $d_{ki}$ in Eq.(\ref{dki}), we have
\begin{eqnarray}\label{n1ds}
n_{1s}=\sum_{i\in c_1} p_i' a_{1i}'d_{1i}\ge \tilde{N}_{1s}=\sum_{i
\in c_1}{1\over 1+\max_{i\in c_1}({p_i a_{1i}\over p'_i
a'_{1i}})} \nonumber \\
n_{1d}=\sum_{i\in c_1} p_i a_{1i}d_{1i}\leq
n_{1d}+n_{1s}-\tilde{N}_{1s}=\max_{i \in c_1}({p_i a_{1i}\over p'_i
a'_{1i}})\tilde{N}_{1s}
\end{eqnarray}
 Eqs.(\ref{popde}, \ref{popse}) can be written in
\begin{eqnarray}\label{d0d}
N_d= n_{0d}+\max_{j\in c_1}({p_j a_{1j}\over p'_j
a'_{1j}})\tilde{N}_{1s} +\Lambda -\xi_1\\
N_s=n_{0s}+ \tilde{N}_{1s} +\Lambda' +\xi_1
\label{s0s}\end{eqnarray} where
\begin{equation}
\Lambda=\sum_{k=2}^J \sum_{i\in c_k} p_i a_{ki}d_{ki};\;
\Lambda'=\sum_{k=2}^J \sum_{i\in c_k}p'_i a'_{ki} d_{ki},
\end{equation}
and
  \begin{equation}
  \xi_1=n_{1s}-\tilde{N}_{1s}\ge 0
\end{equation}
Using the expression of $\Lambda'$, we will have:
\begin{equation}
\Lambda'=\sum_{k=2}^J\sum_{i\in c_k}{p'_i a'_{ki}\over p_i a_{ki}+p'_i a'_{ki}} \geq   \sum_{k=2}^J\sum_{i\in c_k}{1 \over 1+\max_{j\in c_k}({p_j a_{kj}\over p'_j a'_{kj}})}
 \doteq \tilde{\Lambda}
\end{equation}

Further, we assume the important condition
\begin{equation}\label{ace}
\max_{j\in c_k}({p_j a_{kj}\over p'_j a'_{kj}})\leq  \max_{j\in c_2}({p_j a_{2j}\over p'_j a'_{2j}}) \leq \max_{j\in c_1}({p_j a_{1j}\over p'_j a'_{1j}}),\;{\rm for\; all}\;\; k\ge 2.
\end{equation}
So we can write $\Lambda'=\tilde{\Lambda}+\xi_2$ and $\xi_2\ge 0$.
We also have:
\begin{eqnarray}\label{lambda}
\Lambda&=& \sum_{k=2}^J \sum_{i\in c_k}{p_i a_{ki}\over p_i a_{ki}+  p'_i a'_{ki}} \nonumber
\\
&=&\sum_{k=2}^J \sum_{i\in c_k}(1-{p'_i a'_{ki}\over p_i a_{ki}+  p'_i a'_{ki}}) \nonumber
\\
&=&\sum_{k=2}^J \sum_{i\in c_k}(1-{1 \over 1+\max_{j\in c_k}({p_j a_{kj}\over p'_j a'_{kj}})} )-\xi_2 \nonumber
\\
&=&\sum_{k=2}^J \sum_{i\in c_k}{\max_{j\in c_k}({p_j a_{kj}\over p'_j a'_{kj}}) \over 1+\max_{j\in c_k}({p_j a_{kj}\over p'_j a'_{kj}})} -\xi_2\nonumber
\\
&=& \max_{j\in c_2}({p_j a_{2j}\over p'_j a'_{2j}})\tilde{\Lambda}-\xi_2-\xi_3
\end{eqnarray}
where $\xi_3= \sum_{k=2}^J \sum_{i\in c_k}{\max_{j\in c_2}({p_j
a_{2j}\over p'_j a'_{2j}})-\max_{j\in c_k}({p_j a_{kj}\over p'_j
a'_{kj}}) \over 1+\max_{j\in c_k}({p_j a_{kj}\over p'_j
a'_{kj}})}\geq 0$, according to Eq.(\ref{ace}).

With Eq.(\ref{lambda}),  Eq.(\ref{s0s}) is
equivalent to
\begin{eqnarray}
N_d&=&n_{od}+\max_{j\in c_1}({p_j a_{1j}\over p'_j a'_{1j}})\tilde{N}_{1s}+\max_{j\in c_2}({p_j a_{2j}\over p'_j a'_{2j}})
\tilde{\Lambda}-\xi_1-\xi_2-\xi_3 \nonumber \\
N_s&=&n_{0s}+ \tilde{N}_{1s} +\tilde{\Lambda} +\xi_1+\xi_2
\label{s0s1}
\end{eqnarray} Given the Eqs.(\ref{d0d},
\ref{s0s1}),
 we can formulate $\tilde{N}_{1s}$:
\begin{equation}
\tilde{N}_{1s}={N_d-\max_{j \in c_2}({p_j a_{2j}\over p'_j
a'_{2j}})N_s + \max_{j \in c_2}({p_j a_{2j}\over p'_j
a'_{2j}})n_{0s}-n_{0d} +\max_{j \in c_2}({p_j a_{2j}\over p'_j
a'_{2j}})(\xi_1 +\xi_2)+\xi_1+\xi_2+\xi_3\over \max_{j \in c_1}({p_j
a_{1j}\over p'_j a'_{1j}}) -\max_{j \in c_2}({p_j a_{2j}\over p'_j
a'_{2j}})}
\end{equation}
Since $\xi_1,\xi_2,$ and $\xi_3$ are all non-negative, and $\max_{j
\in c_1}({p_j a_{1j}\over p'_j a'_{1j}}) -\max_{j \in c_2}({p_j
a_{2j}\over p'_j a'_{2j}})\ge 0$ by Eq.(\ref{ace}), we now have
\begin{equation}\label{med}
\tilde{N}_{1s}\geq{N_d-\max_{j \in c_2}({p_j a_{2j}\over p'_j
a'_{2j}})N_s + \max_{j \in c_2}({p_j a_{2j}\over p'_j
a'_{2j}})n_{0s}-n_{0d} \over \max_{j \in c_1}({p_j a_{1j}\over p'_j
a'_{1j}}) -\max_{j \in c_2}({p_j a_{2j}\over p'_j a'_{2j}})}
\end{equation}
Therefore, we can now bound the fraction of single counts among all
counts caused by the signal source
\begin{eqnarray}\label{main0}
\Delta_1'\ge \frac{\tilde{N}_{1s}}{N_s}\geq {N_d-\max_{j \in
c_2}({p_j a_{2j}\over p'_j a'_{2j}})N_s + \max_{j \in c_2}({p_j
a_{2j}\over p'_j a'_{2j}})n_{0s}-n_{0d} \over N_s(\max_{j \in
c_1}({p_j a_{1j}\over p'_j a'_{1j}}) -\max_{j \in c_2}({p_j
a_{2j}\over p'_j a'_{2j}}))}\nonumber\\
\geq {N_d-\max_{j}({p_j
a_{2j}\over p'_j a'_{2j}})N_s + \max_{j}({p_j a_{2j}\over p'_j
a'_{2j}})n_{0s}-n_{0d} \over N_s(\max_{j}({p_j a_{1j}\over p'_j
a'_{1j}}) -\max_{j}({p_j a_{2j}\over p'_j a'_{2j}}))}
\end{eqnarray}
The last inequality is due to the fact that if we replace ${\rm
max}_{j\in c_k}$ by ${\rm max}_{j\in C}={\rm max}_{j}$ in all the
previous derivation including Eq.(\ref{ace}), all the intermediate
inequalities still hold. For the decoy source, we can bound the
fraction of single-photon counts by:
\begin{equation}\label{maind}
\Delta_1\geq \min_j({p_j a_{1j} \over p'_j a'_{1j}})\Delta'_1
\end{equation}
where   ${\rm min}_{j}={\rm min}_{j\in C}$.
 In calculating the final key rate, we also need the relation
between quantum bit error rate (QBER) for single-photon counts due
to the decoy source $e_{1d}$ and due to the signal source $e_{1s}$.
Suppose $i\in c_1$ and quantum bit flip probability is $e_{1i}$.
Then the total errors for single photon counts from each sources
are:
\begin{eqnarray}
e_{1d}&=&{\sum_{i\in c_1} {p_i a_{1i} e_{1i}\over p_i
a_{1i}+p'_i a'_{1i} }\over n_{1d}} \\
e_{1s}&=&{\sum_{i\in c_1} {p'_i a'_{1i} e_{1i}\over p_i a_{1i}+p'_i
a'_{1i} }\over n_{1s}}
\end{eqnarray}
Since $n_{1s}=\sum_{i\in c_1} {p'_i a'_{1i} \over p_i a_{1i}+p'_i
a'_{1i} }=\sum_{i\in c_1} {p_i a_{1i} {p'_i a'_{1i}\over p_i
a_{1i}}\over p_i a_{1i}+p'_i a'_{1i} }$, we get:
\begin{equation}
n_{1s}\geq {1 \over \max_j({p_j a_{1j} \over p'_j a'_{1j}})}n_{1d}
\end{equation}
And there is:
\begin{equation}
\sum_{i\in c_1} {p'_i a'_{1i} e_{1i}\over p_i a_{1i}+p'_i a'_{1i}
}\leq {1 \over \min_j({p_j a_{1j} \over p'_j a'_{1j}})}\sum_{i\in
c_1} {p_i a_{1i} e_{1i}\over p_i a_{1i}+p'_i a'_{1i} }
\end{equation}
So:
\begin{equation}
e_{1s}\leq {\max_j({p_j a_{1j} \over p'_j a'_{1j}}) \over
\min_j({p_j a_{1j} \over p'_j a'_{1j}})}e_{1d}
\end{equation}
In the same reason, we can also give a lower bound of $e_{1s}$.
Finally we obtain
\begin{equation}\label{maine}
{\min_{j}({p_j a_{1j} \over p'_j a'_{1j}}) \over \max_{j}({p_j
a_{1j} \over p'_j a'_{1j}})}e_{1d}\leq e_{1s}\leq {\max_{j}({p_j
a_{1j} \over p'_j a'_{1j}}) \over \min_{j}({p_j a_{1j} \over p'_j
a'_{1j}})}e_{1d}
\end{equation}
If ${p_j a_{1j}\over p'_j a'_{1j}}$ and ${p_j a_{2j}\over p'_j
a'_{2j}}$ are constant numbers for all $j\in C$, then
$e_{1s}=e_{1d}$ and the last inequality in Eq.(\ref{main0}) is
identical to Eq.(18) of Ref.\cite{wangyang}, which is the formula of
fraction of single-photon counts with a {\em stable} source.  We
conclude the following theorem:
\\{\bf Theorem:} Decoy-state QKD with a fluctuating source can be
used as if the source were stable if ${p_j a_{1j}\over p'_j
a'_{1j}}$ and ${p_j a_{2j}\over p'_j a'_{2j}}$ are constant numbers
for all $j\in C $.

Interestingly, the condition in the theorem is {\em not} equivalent
to Eq.(\ref{elem}), the elementary proposition given by
Hwang\cite{H03}. Our theorem only depends on the fluctuation
characterization of those pulses in set $C$, i.e., those pulses
accompanied by a click of Bob's detector, while Eq.(\ref{elem})
actually requests $n_{kd}/N_{kd}=n_{1d}/N_{1d}$ therefore  needs the
fluctuation characterization of {\em all} pulses from Alice's lab.
Here $N_{kd}$ and $N_{ks}$ are number of k-photon pulses from decoy
source and that from signal source, respectively. Given this fact,
one can in principle construct specific cases where Eq.(\ref{elem})
does {\em not} hold while the condition in our theorem still holds.
  This shows that our Eq.(\ref{popdk}) is indeed
more fundamental and more general than Eq.(\ref{elem}).

\subsection{Conclusion on AYKI protocol}
 The AYKI protocol uses a heralded source of:
\begin{equation}\label{psi}
\ket{\Psi_i}_{AS}=\sum_{k=0}^\infty \sqrt{X_{ki}} \ket{k}_A\ket{k}_S
\end{equation}
where $X_{ki}=\mu_i^k (1+\mu_i)^{-(k+1)}$. The heralded source state
can be produced by, e.g., the parametric down conversion (PDC) which
is pumped by strong light pulses whose intensity fluctuation can be
as large as 20\%. In the protocol, Alice detects mode $A$ and sends
out mode $S$ to Bob. Mode $S$ is a decoy pulse when Alice's detector
clicks and a signal pulse when her detectors does not click. Suppose
Alice's detector has a detecting efficiency $\eta_A$ and dark count
rate $d_A$.  At any time $i$, given the two mode state in
Eq.(\ref{psi}), the probability that her detector clicks or not is
\begin{eqnarray}p'_i={d_A+\mu_i \eta_A\over 1+\mu_i \eta_A}\nonumber\\p_i={1-d_A
\over 1+\mu_i \eta_A }\end{eqnarray} When the detector clicks or
not, we obtain a signal state or a decoy state  at mode $S$  in the
form of Eq.(\ref{rhoi}), and
\begin{eqnarray}
a_{ki}&=&{1+\mu_i \eta_A\over 1-d_A}X_{ki} (1-\gamma_k) \nonumber \\
a'_{ki}&=&{1+\mu_i \eta_A \over d_A+\mu_i \eta_A}X_{ki} \gamma_k
\end{eqnarray}
and
\begin{equation}
\gamma_k=1-(1-d_A)(1-\eta_A)^k
\end{equation}
We find
\begin{equation}
{p_i a_{ki} \over p'_i a'_{ki}}={1-\gamma_k \over
\gamma_n}={(1-d_A)(1-\eta_A)^k\over 1-(1-d_A)(1-\eta_A)^k}
\end{equation}
They are {\em independent} on $i$, hence source fluctuation does not
change the final formula of the protocol by our theorem. Also,
according to Eq.(\ref{main0}) we can write:
\begin{equation}
n_{1s}= \tilde{N}_{1s}\geq{N_d - {1-\gamma_2 \over
\gamma_2}N_s+{1-\gamma_2 \over \gamma_2}n_{0s}-n_{0d}\over
{1-\gamma_1 \over \gamma_1}-{1-\gamma_2 \over \gamma_2}}
\end{equation}
Using the fact  ${n_{0s}\over n_{0d}}={p'_i a'_{0i}\over p_i
a_{0i}}={d_A \over 1-d_A}$ we have
\begin{equation}
\Delta'_1\geq{N_d - {1-\gamma_2 \over
\gamma_2}N_s-(1-{1-\gamma_2\over \gamma_2}{d_A \over
1-d_A})n_{0d}\over N_s({1-\gamma_1 \over \gamma_1}-{1-\gamma_2 \over
\gamma_2})}
\end{equation}
 which is just the major formula in
Ref.\cite{AYKI}. We have {\em not} used Eq.(\ref{elem}) in our
proof, though the derivation given by AYKI\cite{AYKI} have assumed
one constant intensity $\mu$ and used the assumption of
Eq.(\ref{elem}). Our proof concludes that the AYKI protocol actually
works securely with whatever intensity fluctuation, though the key
rate can be low if the source fluctuation is large, in which case
one may observe poor values of $N_s$ and $N_d$.
\section{Improved formula for 3-intensity protocol}
If we add a vacuum source to the 2-intensity protocol and uses
vacuum source with probability $p_{vi}$ at the $i$th time, we have a
3-intensity protocol where one can estimate number of vacuum counts
$n_{0s}$, $n_{0d}$ more precisely therefore improve the final key
rate. Suppose the vacuum source caused $N_0$ clicks. By similar
derivation done in Ref.\cite{wangyang}, we obtain
\begin{equation}\label{dk}
n_{0d}\leq {1\over p_0}\max_{j \in \tilde c_0}(p_j a_{0j})\sum_{i\in
\tilde c_0}p_0 d_{0i}\leq {1\over p_0}\max_{j}(p_j a_{0j})N_0.
\end{equation}
\subsection{Numerical simulation} In our major formula,
Eq.(\ref{main0}), there are terms of  ${\rm max}_i$, this {\em
economic} worst-case estimation can be significantly smaller than
the {\em normal} worst-case of $\frac{{\rm max}_i(p_i a_{ki})}{{\rm
min}_i(p'_ia'_{ki})}$ estimation as proposed in Ref.\cite{wangyang},
hence Eq.(\ref{main0}) can improve the key rate {\em a lot}.
Consider the model that both decoy pulse and signal pulse are
generated through attenuating a common father pulse at time $i$. We
set $p_{vi}=p_0$, $p_i=p$ and $p_i'=p'$ to be constants. The
fluctuation of the final pulse out of Alice's lab  consists of two
parts: father pulse fluctuation and device (attenuator) parameter
fluctuation. Suppose Alice {\em wants} to use intensity $\mu,\;\mu'$
for her decoy pulse and signal pulse. She wants to obtain them
through attenuating the father pulse of intensity $F$ by setting her
attenuator's transmittance to be $\lambda_D =
\frac{\mu}{F},\;\lambda_S = \frac{\mu'}{F}$ for a decoy pulse or for
a signal pulse. However, the actual case is that at any time $i$,
the intensity of the father pulse is $F_i = F(1+\delta_i)$,
$\lambda_{Di}=\lambda_D(1+\epsilon_{id})$ and
$\lambda_{Si}=\lambda_S(1+\epsilon_{is})$. We have the upper bounds
of: $|\delta_i|\le \delta$, $|\epsilon_{id}|\le \epsilon_d$ and
$|\epsilon_{is}|\le \epsilon_s$. The actual intensity of the  $i$th
pulse out of Alice's lab is
\begin{eqnarray}
{\rm decoy:}\;\;\;\mu_{i}&=&\mu (1+\delta_i)(1+\epsilon_{id}) \\
{\rm signal:}\;\;\;\mu_{i}'&=&\mu' (1+\delta_i)(1+\epsilon_{is}) \\
\max_i({p a_{ki}\over p' a'_{ki}})&=&{p\over p'}({\mu(1+\epsilon_d )
\over
\mu'(1-\epsilon_s)})^k\exp\{(1+\delta)[\mu'(1-\epsilon_s)-\mu(1+\epsilon_d)]\}
\\
& &for\;k\geq 1
\end{eqnarray}
In a real experiment, only {one} pulse is prepared and sent out at
any $i$th time. Values $\mu_{i}$ and $\mu_{i}'$ can be interpreted
as the would-be values if Alice decided to produce a decoy pulse or
signal pulse at the $i$th time, given a certain set-up. The bound of
vacuum count in Eq.(\ref{dk}) is now $n_{0d}\leq {p N_0 \over
p_0}e^{-{\rm min}_i(\mu_{i})}$ and $n_{0s}\geq {p' N_0 \over
p_0}e^{-{\rm max}_i(\mu_{i}')}.$
 Note that here we
have taken the normal worst-case estimation only for the device
fluctuation, while we have taken the {\em economic} worst-case
estimation for the intensity fluctuation of the father pulse.  Our
result here also applies to a source state a little bit different
from a coherent state: we just add very small new fluctuation terms
to the parameters $a_{ki},\;a_{ki}'$ in the states. If these terms
are negligibly small, there effects to the final key rate is also
negligible.
 Though
there are also other methods\cite{wangapl,yi,yilo} for decoy-state
QKD with fluctuating source, they\cite{wangapl,yi,yilo} only apply
to the father pulse fluctuation, but not apply to the device
fluctuation, as clearly pointed out in\cite{yi,njp}. As shown
already, our method does not assume zero device fluctuation since it
directly applies to the fluctuation of the final pulse out of
Alice's lab.

 We use the experimental
data for 50km given by done by Peng et al \cite{peng} QKD for
numerical simulation. The results are shown in Fig.1, where we set
$\epsilon_d=\epsilon_s=\epsilon$. We find that the fluctuation of
the father pulse intensity changes the final key rate {\em very}
slightly using Eq.(\ref{main0}) in this work in the calculation,
while the device fluctuation still degrades the final key rate
drastically. Our results here also apply to the Plug-and-Play
protocol\cite{gisind}.
\begin{figure}
\includegraphics[width=200pt]{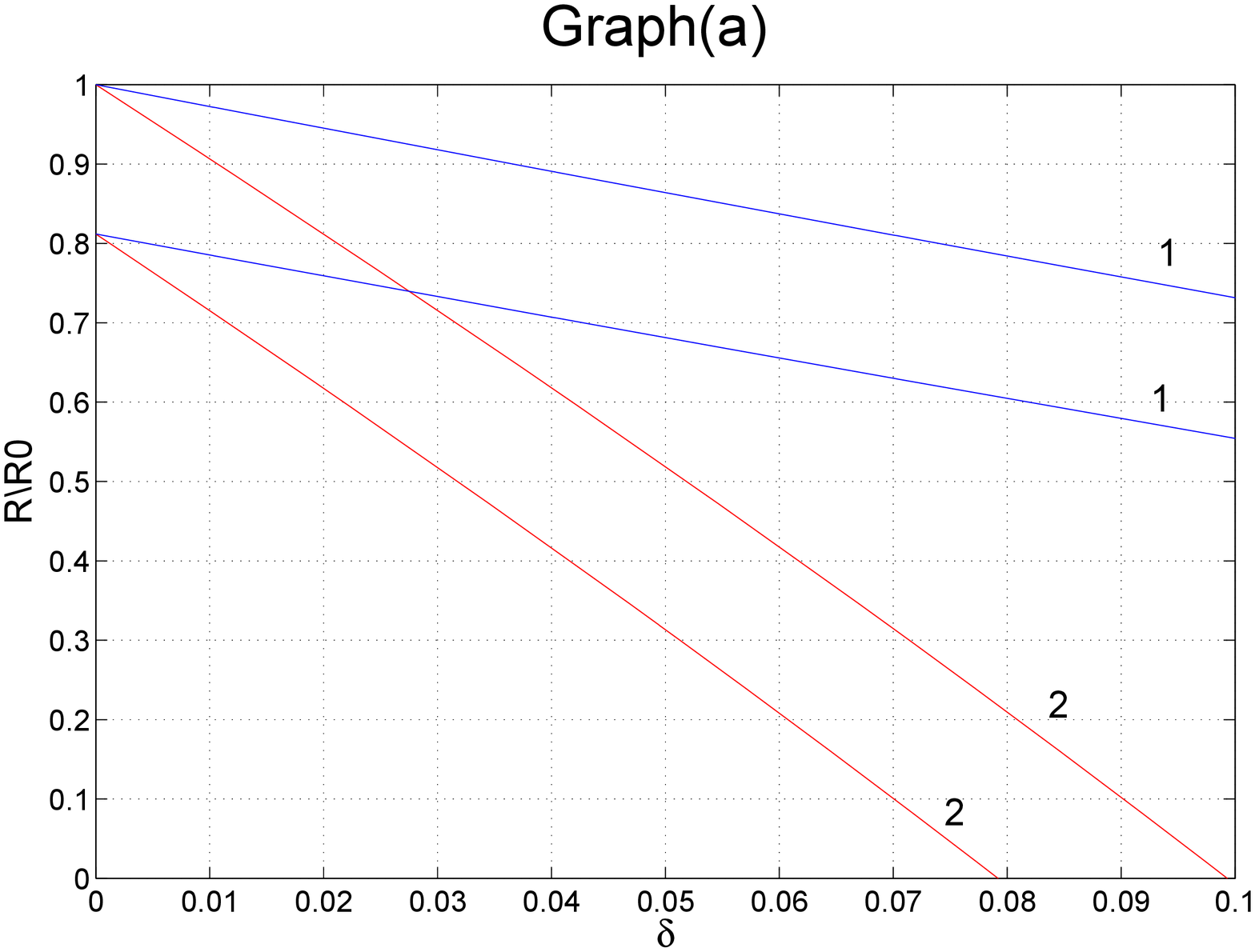}
  \includegraphics[width=190pt]{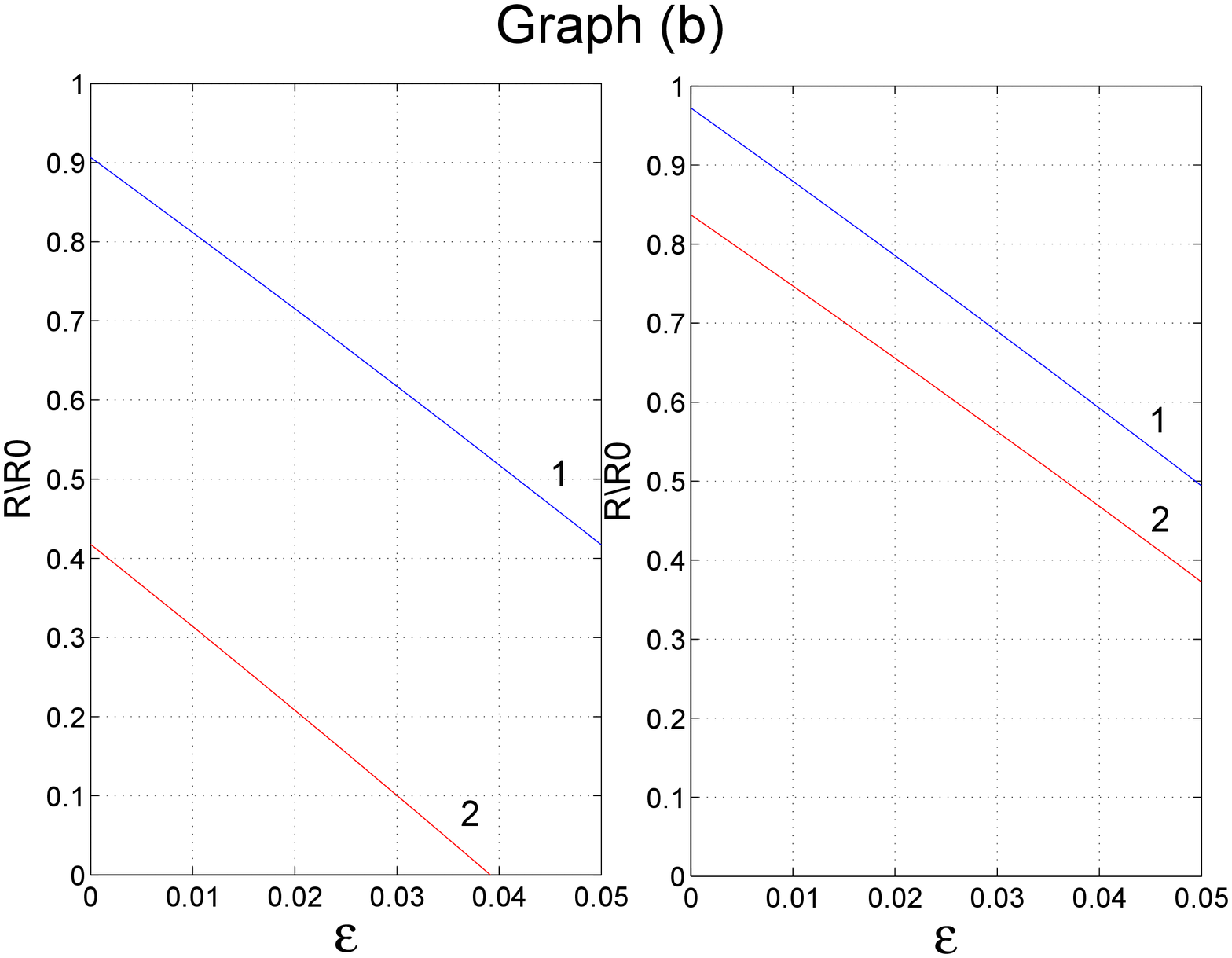}\\
  \caption{Comparison of relative key rates to that of zero fluctuation.
  $a$: Relative key rates with father-pulse-intensity fluctuation
  at $\epsilon=0$ and $2\%$. Line 1 and 2 are calculated with Eq.(\ref{main0}) of this work
  and Eq.(58) of Ref.\cite{wangyang}, respectively.  $b$: Relative key rate with device fluctuation.
  $b$(left):  Result calculated with Eq.(58)
  of Ref.\cite{wangyang}; $b$(right) Result calculated with
   Eq.[\ref{main0}]. Line 1 and 2 in $b$ are for
$\delta=1\%,6\% $, respectively. }
\end{figure}
\section{ Concluding remark and discussions} In summary, we have presented
a criteria for the secure decoy-state protocol with source
fluctuation. The AYKI protocol satisfies the criteria. A well known
advantage of AYKI protocol is that there is no need to make
intensity switching. Given our proof here, AYKI protocol neither
needs intensity switching nor needs source intensity monitoring. As
the base of the conclusion, our general formulas here are efficient
for other protocols, such as the 3-intensity protocol using coherent
states. It should be interesting to include the finite size
effects\cite{ILM,haya,cai} in the future study.
 \\{\bf Acknowledgement:} We thank L. Yang for his help in the numerical
 simulation. This work was supported in part by the National Basic Research
Program of China grant No. 2007CB907900 and 2007CB807901, NSFC grant
No. 60725416 and China Hi-Tech program grant No. 2006AA01Z420.


\begin{thebibliography}{99}
\bibitem{BB84}
C.H.~Bennett and
G.~Brassard, in {\em Proc.\ of IEEE Int.\ Conf.\ on Computers,
Systems, and Signal Processing (IEEE, New York, 1984)},
pp.~175-179.
\bibitem{GRTZ02}
N.~Gisin, G.~Ribordy, W.~Tittel, and H.~Zbinden, Rev. Mod. Phys.
{\bf 74}, 145 (2002); N. Gisin and R. Thew, Nature Photonics, 1, 165
(2006); M.~Dusek, N.~L\"utkenhaus, M.~Hendrych, in {\em Progress in
Optics VVVX}, edited by E.~Wolf (Elsevier, 2006); V. Scarani, H.
Bechmann-Pasqunucci, N.J. Cerf, M. Dusek, N. L\"utkenhaus, and M
Peev, Rev. Mod. Phys. 81, 1301 (2009).
\bibitem{ILM}
H.~Inamori, N.~L\"utkenhaus, D.~Mayers, European Physical Journal D,
41, 599 (2007), which appeared in the arXiv as quant-ph/0107017.
\bibitem{PNS}
B.~Huttner, N.~Imoto, N.~Gisin, and T.~Mor, Phys. Rev. A {\bf 51},
1863 (1995); H.P.~Yuen, Quantum Semiclassic. Opt. {\bf 8}, 939 (1996).
\bibitem{PNS1}
G.~Brassard, N.~L\"utkenhaus, T.~Mor, and
B.C.~Sanders, Phys. Rev. Lett. {\bf 85}, 1330 (2000);
N.~L\"utkenhaus, Phys. Rev. A {\bf 61}, 052304 (2000);
N.~L\"utkenhaus and M.~Jahma, New J. Phys. {\bf 4}, 44 (2002).
\bibitem{gllp}
D.~Gottesman, H.K.~Lo, N.~L\"{u}tkenhaus, and J.~Preskill, Quantum
Inf. Comput. {\bf 4}, 325 (2004).
\bibitem{rep}X.-B. Wang, T. Hiroshima, A. Tomita, and M. Hayashi,
{\em Physics Reports} 448, 1(2007)
\bibitem{H03}
W.-Y.~Hwang, Phys. Rev. Lett. {\bf 91}, 057901 (2003).

\bibitem{wang05}
X.-B.~Wang, Phys. Rev. Lett. {\bf 94}, 230503 (2005); X.-B.~Wang,
Phys. Rev. A {\bf 72}, 012322 (2005).


\bibitem{LMC05}
H.-K.~Lo, X.~Ma, and K.~Chen, Phys. Rev. Lett. {\bf 94}, 230504
(2005); X.~Ma, B. Qi, Y. Zhao, and H.-K. Lo, Phys. Rev. A {\bf
72}, 012326 (2005).

\bibitem{HQph}
J.W.~Harrington {\em et al.}, quant-ph/0503002.
\bibitem{AYKI}Y. Adachi, T. Yamamoto, M. Koashi, and N. Imoto, Phys. Rev.Lett.
{\bf 99}, 180503 (2007).
\bibitem{haya}M. Hayashi, Phys. Rev. A 74, 022306 (2008); ibid 76,
012329 (2007).

\bibitem{scran}
V.~Scarani, A.~Acin, G.~Ribordy, N.~Gisin, Phys. Rev. Lett. 92,
057901 (2004); C.~Branciard, N.~Gisin, B.~Kraus, V.~Scarani, Phys.
Rev. A 72, 032301 (2005).

\bibitem{kko} M. Koashi, Phys. Rev. Lett., 93, 120501(2004);
K.~Tamaki, N.~L\"ukenhaus, M.~Koashi, J.~Batuwantudawe,
quant-ph/0608082

\bibitem{peng}   D. Rosenberg {em et al.},  {\em Phys. Rev. Lett.} 98, 010503
(2007);  T. Schmitt-Manderbach {\em et al.}, {\em Phys. Rev. Lett.}
98, 010504 (2007); Cheng-Zhi Peng {\em et al.}
 Phys. Rev. Lett. 98, 010505 (2007); Z.-L. Yuan, A. W. Sharpe, and A. J. Shields, {\em
Appl. Phys. Lett.} 90, 011118 (2007); Y.~Zhao, B. Qi, X. Ma, H.-K.
Lo and L. Qian, Phys. Rev. Lett. {\bf 96}, 070502 (2006); Y. Zhao,
B. Qi, X. Ma, H.-K. Lo, and L. Qian, in Proceedings of IEEE
International Symposium on Information Theory, Seattle, 2006, pp.
2094--2098 (IEEE, New York).
\bibitem{wang07}X.-B. Wang, {\em Phys. Rev.} A75, 012301(2007)
\bibitem{wangapl}X.-B. Wang, C.-Z. Peng and J.-W. Pan, {\em Appl. Phys.
Let.} 90, 031110(2007)
\bibitem{wangyang} X.-B. Wang, C.-Z. Peng {\em et al.} {\em Phys. Rev.
A} 77, 042311 (2008).
\bibitem{yi}Y. Zhao {\em et al}, Phys. Rev. A {\bf 77}, 052327 (2008).
\bibitem{njp}X.-B. Wang, L. Yang, C.-Z. Peng and J.-W. Pan, New J. Phys. 11, 075006
(2009).
\bibitem{yilo} Y. Zhao, B. Qi, H.-K. Lo, and Q. Li, New J. Phys. 12, 023024 (2010).
\bibitem{guo1}F.X. Xu, Y. Zhang, Z. Zhou, W. Chen, Z. F. Han, and G.
C. Guo, Phys. Rev. A, 80, 062309 (2009).
\bibitem{guo2}X. Peng, H. Jiang, B.J. Xu, X.F. Ma, and H. Guo, Opt.
Lett., 33, 2077 (2008).
\bibitem{gisind}N. Gisin, S. Fasel, B. Kraus, H. Zbinden, and G.
Ribordy, Phys. Rev. A 73, 022320(2006).
\bibitem{cai}Raymond Y. Q. Cai and V. Scarani, New J. Phys., 11, 11, 045024
(2009).
\end{thebibliography}
\end{document}